\documentclass[12pt]{iopart}

\usepackage{cite}
\usepackage{dcolumn}
\usepackage{bm}
 \expandafter\let\csname equation*\endcsname\relax
  \expandafter\let\csname endequation*\endcsname\relax
\usepackage{amsmath}
\usepackage{bm,braket}
\usepackage{color}

\usepackage{float}
\usepackage{graphicx}

\begin{document}

\abovedisplayskip=8pt
\abovedisplayshortskip=8pt
\belowdisplayskip=8pt
\belowdisplayshortskip=8pt

\title[Phonon-mediated population inversion in a semiconductor QD cavity system]{Phonon-mediated population inversion\\ in a semiconductor quantum-dot cavity system}

\author{S.~Hughes}
\address{Department of Physics,
Queen's University,
 Kingston, Ontario,  Canada K7L 3N6}
\ead{shughes@physics.queensu.ca}
\author{H.~J.~Carmichael}
\address{Department of Physics, University of Auckland, Private Bag 92019, Auckland,\\ New Zealand}
\ead{h.carmichael@auckland.ac.nz}

\begin{abstract}

We investigate pump-induced exciton inversion in a quantum-dot cavity system with continuous wave drive. Using a polaron-based master equation, we demonstrate excited-state populations above $0.9$ for an InAs dot at a phonon bath temperature of $4\mkern2mu{\rm K}$. In an exciton-driven system, the  dominant mechanism is incoherent excitation from the phonon bath.
For cavity driving, the mechanism is phonon-mediated switching between ground- and excited-state branches of the ladder of photon states, as quantum trajectory simulations clearly show. The exciton inversion as a function of detuning is found to be qualitatively different for exciton and cavity driving, primarily due to cavity filtering. The master equation approach allows us to include important radiative and non-radiative decay processes on the zero phonon line, provides a clear underlying dynamic in terms of photon and phonon scattering, and admits simple analytical approximations that help to explain the physics.


\end{abstract}


\maketitle


\section{Introduction}
The ability to achieve inversion and lasing in atomic and solid state systems is a topic of continuing interest \cite{Eberly:book}. In the domain of cavity-QED, regimes of ``single atom lasing'' and non-classical light emission have been studied \cite{Mu:PRA92,McKeever:Nature03}.

According to the common wisdom, any scheme to achieve population inversion must exploit a multi-level manifold of material states; steady-state inversion is not possible for a driven two-level system. This wisdom is restricted, however, to a simple two-level system (TLS) coupled to classical fields in a non-engineered environment. When interacting with a \emph{quantized} radiation mode, a TLS may be inverted by two-photon excitation to higher lying levels of the  quantized \emph{matter plus field} \cite{Savage:PRL88,Hughes:PRL11,Leek:PRB09}. Also, when in interaction with an environment near a photonic bandgap, a TLS may be inverted by way of an asymmetry of the Mollow triplet; although, very high (many orders of magnitude) contrast in the sideband decay rates is required  \cite{John:PRA01}.

In contrast to atoms, the unique features of semiconductor quantum dots (QDs)
include their large optical dipole moment and engineerable emission wavelength. Other attractive features are their fixed position and potential for  integration with cavities and waveguides using developed semiconductor fabrication techniques \cite{Hennessy:Nature07}.
Although semiconductor microcavity lasers are today quite common \cite{Jewell:APL89,Painter:Science99,Vahala:Nature03,Vuckovic:Nature11},
there is an ongoing push to realize more exotic regimes, such as ``single QD lasing,'' with the aim of producing a non-classical light source in a solid state nanostructure. Single dot lasing \emph{per se} is  not addressed in this work. Rather, we focus on the fundamental mechanisms available to achieve population inversion in a semiconductor QD-cavity system.

Semiconductor QDs differ from atoms at the level of fundamental physics, and this calls for extra care when describing the light-matter interaction. Typically, QDs are embedded in a solid state lattice where electron-phonon interactions, though sometimes ignored in quantum optical studies, are known to impact optical properties; they affect photoluminescence lineshapes \cite{Besombes:PRB01}, coherent Rabi oscillation \cite{Ramsay:PRL10}, and the Mollow triplet spectrum of resonance fluorescence \cite{Ulrich:PRL11,Roy&Hughes:PRL11}; phonon-mediated scattering can cause excitation-induced dephasing \cite{Knorr:PRL03,Ramsay:PRL10}, which is detrimental to the exploitation of quantum optical interactions. It is interesting, then, to ask what impact this scattering has when one tries to coherently drive an excitonic transition coupled to a quantized radiation mode into a regime of population inversion \cite{Savage:PRL88,Hughes:PRL11,Leek:PRB09}. In answer we find a surprising result: contrary to the expectation that it might be detrimental, acoustic-phonon-mediated scattering can be exploited to achieve significant exciton inversion via a mechanism entirely different to those previously reported to invert a simple TLS. The mechanism is richer and significantly more efficient. It relies on the natural asymmetry of phonon emission and absorption at low bath temperatures, and for cavity driving, exploits a one-way coupling (phonon-assisted scattering) between the exciton and the cavity mode when the latter is blue detuned with respect to the zero phonon line (ZPL).

Population inversion in a QD in the presence of phonons without cavity coupling has been reported in previous theoretical studies. Gl\"assl {\em et al.} \cite{Glassl:PRB11} employed path integral and correlation expansion techniques to explore the long-time behaviour of a coherently driven QD with phonon coupling and noted that away from resonance large population inversions can be realized. The authors compare their results with a simple model based on thermalization
in dressed states and find good agreement for the parameters considered. One limitation of this work is the omission of ZPL decay processes; these are generally needed for the quantitative modeling of experiments. Stace {\em et al.}\cite{Tom:PRL05} also studied the driven exciton-phonon system, again with no cavity coupling, and report population inversion in a {\em unstructured} phonon bath. They consider a coherently driven double-dot system in the large pump limit in a dressed-state approach. Phonon coupling is included as a perturbation and exciton inversion reported for a phonon bath with subquadratic spectral density.

Both of these works exploit phonons to achieve TLS inversion with continuous wave (cw) drive. With the QD embedded in a cavity, it is not known how these processes change, nor is it known what differences, if any, arise between the two methods of driving now made available: direct driving of the exciton or driving through the cavity. It is not known how ZPL decay processes effect the achievable population inversion.

\begin{figure}[t]
\centering
\includegraphics[width=0.8 \columnwidth]{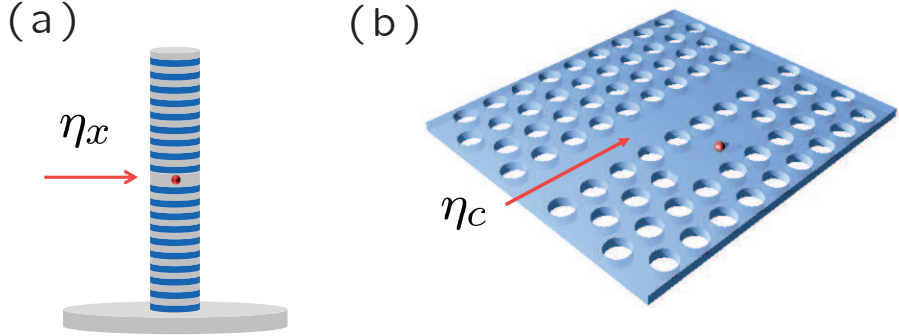}
\vspace{-0.0cm}
\caption{(Color online) Schematics of two possible semiconductor cavity-QED systems with coherent exciton ($\eta_x$) or cavity ($\eta_c$) driving: (a) a micropillar system and (b) a planar photonic crystal system. A small red sphere shows the position of the QD.}
\label{fig:schematic1}
\end{figure}

In this work we investigate several mechanisms realizing pump-induced exciton inversion in a QD-cavity system with cw driving. We propose and characterize the phonon-assisted scheme using a semiconductor quantum-optics formalism that includes phonon interactions by way of an effective master equation (ME) \cite{Roy&Hughes:PRX11}. The ME uses a polaron transform \cite{Roy&Hughes:PRX11,Roy&Hughes:PRL11} to take electron-acoustic-phonon interactions into account at a microscopic level, and includes cavity and QD decay in a Lindblad description. It yields an intuitive and transparent modeling of the QD-cavity dynamics through a combination of photon and phonon scattering. We consider a QD embedded within a small high $Q$ dielectric cavity, as shown schematically in Figs.~\ref{fig:schematic1}(a) and \ref{fig:schematic1}(b). Either the exciton or cavity mode is subject to coherent cw driving ($\eta_x$ and $\eta_c$ in the figure). A possible advantage of cavity over exciton driving is that it might mitigate problems with excitation-induced dephasing, a process that accompanies excitation via the exciton-phonon reservoir \cite{Ulrich:PRL11,Roy&Hughes:PRL11,Roy&Hughes:PRX11}; it also allows for pumping through a waveguide input channel \cite{Hughes:PRB04} [cf.~Fig.~\ref{fig:schematic1}(b)], allowing chip-based quantum optics using semiconductor fabrication techniques.

The paper is organized as follows. In Sec.~\ref{introduction} we review the ME theory and present a simple analytical solution for the exciton density in the absence of cavity coupling. We also discuss our numerical scheme for treating the full ladder of cavity photons in the presence of either exciton or cavity driving.  Our results for the exciton inversion in a QD-cavity system are presented in Sec.~\ref{results}, where we explore the effect of cavity-exciton detuning and study the difference between exciton versus cavity driving, demonstrating the role of phonon-induced scattering in each case. The path to inversion is found to be qualitatively different for exciton versus cavity driving, with inversion occurring in the latter case only with the cavity suitably blue-shifted with respect to the exciton resonance. We introduce a simple analytical model to explain our results. Conclusions are offered in Sec.~\ref{conclusions}.

\section{Theory}
\label{introduction}
We work in a frame rotating at the laser pump frequency $\omega_L$. The simplest Hamiltonian for a single QD interacting with a cavity mode and phonons, excluding QD and cavity decay, is
\begin{align}
\label{sec1eq1}
H&=\hbar\Delta_{xL}{\sigma}^{+}{\sigma}^{-}+\hbar\Delta_{cL}{a}^{\dagger}{a}+\hbar g({\sigma}^{+}{a}+{a}^{\dagger}{\sigma}^{-})+H^{}_{\rm{drive}}\nonumber\\
&+{\sigma}^{+}{\sigma}^{-}\sum_{q}\hbar\lambda_{q}({b}_{q}+{b}_{q}^{\dagger})+\sum_{q}\hbar\omega_{q}{b}_{q}^{\dagger}{b}_{q}\, ,
\end{align}
where ${b}_{q}^{\dagger}$ and ${b}_{q}$ are creation and annihilation operators for mode $q$ of the phonon bath, $\lambda_q$ is the (real) exciton-phonon coupling, $a^\dagger$ and $a$ are photon creation and annihilation operators for the cavity mode, and ${\sigma}^+$ and ${\sigma}^-$ are Pauli raising and lowering operators for the exicton; $\Delta_{\alpha L}\equiv \omega_\alpha-\omega_L$ ($\alpha =x,c$) designates the detuning of the exciton (frequency $\omega_{x}$) and cavity (frequency $\omega_{c}$)  from the laser drive, and $H_{\rm{drive}}=H^c_{\rm drive}+H^x_{\rm drive} = \hbar\eta_{c}({a}
+{a}^{\dagger})+\hbar\eta_{x}({\sigma^+}+\sigma^-)$ is the drive Hamiltonian. Since we deal with quasi-resonant coherent excitation, higher lying exciton states and continuum levels in the QD material are neglected. We allow for driving of the exciton and the cavity but consider the two cases separately---i.e., we adopt either non-zero $\eta_x$ or non-zero $\eta_c$.

In moving from Hamiltonian (\ref{sec1eq1}) to the effective ME, one first transforms to a polaron frame, which formally recovers the independent boson model \cite{Mahan:Book90,Imamoglu:PRB02,Krummheuer:PRB02} in the appropriate limit; the independent boson model is known to capture the characteristic spectrum of an exciton coupled to a phonon bath \cite{Besombes:PRB01}. The derived ME \cite{Imamoglu:PRB02,Nazir:NJP10,Roy&Hughes:PRL11} treats the coherent electron-phonon interaction nonperturbatively through a mean phonon displacement,
\begin{equation}
\langle B\rangle=\exp\left[-\frac{1}{2} \sum_q\left(\frac{\lambda_q}{\omega_q}\right)^2 (2\bar n_q+1) \right ]
=\exp\left[-\frac{1}{2}\int^{\infty}_{0}d\omega\frac{J(\omega)}{\omega^{2}}\coth\left (\frac{\hbar\omega} {2k_bT} \right )\right],
\end{equation}
where $\bar n_q$ is the mean phonon number (Bose-Einstein distribution at temperature $T$), and $J(\omega) = \sum_q \lambda_q^2\delta(\omega-\omega_q)$ is the phonon spectral function. The incoherent interaction (scattering) is treated within the second-order Born approximation (see Refs.~\cite{Roy&Hughes:PRX11,Roy&Hughes:PRL11} for details), though it is important to note that higher-order contributions are included by the polaron transform. Adding QD and cavity decay, the time-convolutionless ME for the reduced density operator is
\begin{equation}
\label{ME}
\frac{\partial \rho}{\partial t}=\frac{1}{i\hbar}[H_{S}^{\prime},\rho]+{\cal L}_{\rm ph}\rho+{\cal L}\rho,
\end{equation}
with polaron-transformed Hamiltonian
\begin{equation}
H^{\prime}_{S}= \hbar(\Delta_{xL}-\Delta_{P}){\sigma}^{+}{\sigma}^{-}
+\hbar\Delta_{cL} {a}^{\dagger}{a}+\hbar {X}_{g}+H^c_{\rm drive},
\label{eq:H_S}
\end{equation}
polaron shift $\Delta_P=\int^{\infty}_{0}d\omega{J(\omega)}/{\omega}$ (absorbed by $\Delta_{xL}$ below), and
phonon scattering term
\begin{equation}
{\cal L}_{\rm ph}\rho=-\sum_{m=g,u}\int^{\infty}_{0}\mkern-3mud\tau G_{m}(\tau)[{X}_{m},X_m(\tau)\rho]
+{\rm H.c.},
\label{eq:FullRates}
\end{equation}
where $X_m(\tau)=\exp(-iH_{S}^{\prime}\tau/\hbar){X}_{m}\exp(iH_{S}^{\prime}\tau/\hbar)$, with $(X_g,iX_u)=g'({a}^{\dagger}{\sigma}^{-}\pm{\sigma}^{+}{a})+\eta_x'(\sigma^-\pm\sigma^+)$, $g^\prime=\braket{B}g$, $\eta_x^\prime=\braket{B}\eta_x$.
The rescaling $g\rightarrow g'=\braket{B}g$ was pointed out some time ago by Wilson-Rae and Imamo\ifmmode \breve{g}\else \u{g}\fi{}lu \cite{Imamoglu:PRB02}. It is important to note that $g^\prime$ and $\eta_x^\prime$ are temperature dependent; although this dependence is often ignored when fitting experiments, and compensated for by changing other parameters in an attempt to improve the fit. The response functions $G_{g}(t)=\cosh[\phi(t)]-1$ and $G_{u}(t)=\sinh[\phi(t)]$ are polaron Green functions \cite{Mahan:Book90,Imamoglu:PRB02}\footnote{Obtained by assuming a separable density operator for the system and phonon-bath and tracing over the phonon degrees of freedom.}, defined by the phonon phase term
\begin{align}
\label{eq:phase}
\phi(t)& =\sum_q \left(\frac{\lambda_q}{\omega_q}\right)^2  \left [ (\bar n_q+1)e^{-i\omega_q t} + \bar n_q e^{i\omega_q t}  \right ]  \\
& =\int^{\infty}_{0}d\omega\frac{J(\omega)}{\omega^{2}}\left[\coth\left(\frac{\hbar\omega}{2k_bT}\right)\cos(\omega t)-i\sin(\omega t)\right],
\end{align}
which clearly includes contributions from multi-phonon scattering. The last term in Eq.~(\ref{ME}) is a sum of three Lindblads: ${\cal L}=\kappa{\cal L}[a]+(\gamma/2){\cal L}[\sigma^-]+(\gamma^\prime/2){\cal L}[\sigma_{ee}]$, with ${\cal L}[\xi]\rho=2\xi\rho\xi^\dagger-\xi^\dagger\xi\rho-\rho\xi^\dagger\xi$ and $\sigma_{ee}=|e\rangle\langle e|$. It accounts for cavity decay at rate $2\kappa$, exciton decay at rate $\gamma$, and pure dephasing of the exciton at rate $\gamma^\prime$. These processes broaden the ZPL, an essential effect not captured by the independent boson model.

We adopt the established phonon spectral function, $J(\omega)=\alpha_{p}\,\omega^{3}\exp (-{\omega^{2}}/{2\omega_{b}^{2}})$, which describes the electron-acoustic-phonon interaction via a deformation potential, the dominant source of phonon scattering for InAs and GasAs QDs \cite{Takagahara:PRB99}. Appropriate numbers  are obtained by fitting experimental measurements made on InAs QDs \cite{Hughes:PRB11,StuttgartExps1}.

\begin{figure}[t]
\centering
\includegraphics[width=0.8 \columnwidth]{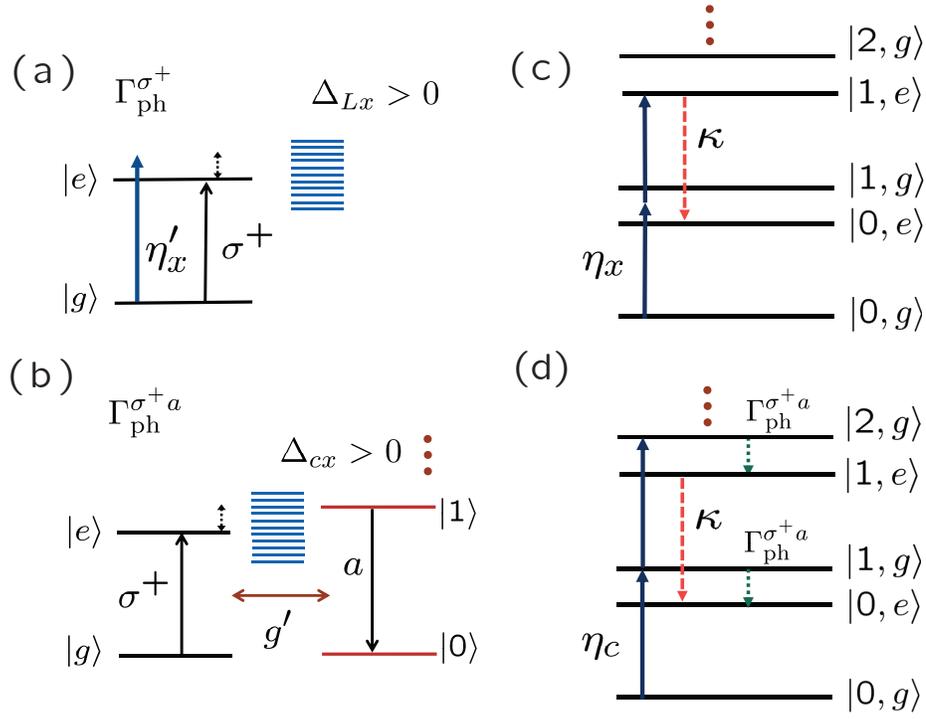}
\vspace{-0.0cm}
\caption{(Color online)
(a) Exciton energy levels ($\ket{e}$ and $\ket{g}$) with exciton coupling to a phonon bath;  the processes, $\eta'_x$  and $\Gamma_{\rm ph}^{\sigma^{+}}$, that result in incoherent excitation of the exciton are shown. (b) Similar to (a) but with coupling to a quantized cavity mode; the first two levels of the cavity ladder ($|0\rangle$ and $|1\rangle$) are shown, along with the phonon-induced mechanism,  $\Gamma_{\rm ph}^{\sigma^{+}a}$, that results in exciton excitation as a cavity photon is emitted; $\Delta_{cx}$ is the cavity-exciton detuning. (c) Energy level diagram showing how exciton inversion can be realized via two-photon pumping of the state $\ket{e,1}$, which then loads state $\ket{0,e}$ via fast cavity emission at rate $2\kappa$; no phonon scattering processes are present.
(d) Simplified level scheme showing the approximate system behavior of ground- and excited-state harmonic oscillators with one-way coupling through phonon-mediated scattering (coherent coupling $g'=\langle B\rangle g$ neglected). }
\label{fig:schematic2}
\end{figure}

Our model as outlined yields an involved solution scheme and little physical insight. We turn therefore to an effective Lindblad form of the phonon scattering term, which is shown by Roy and Hughes \cite{Roy&Hughes:PRX11} to be in very good agreement with the predictions of Eq.~(\ref{ME}). For cavity excitation it makes the replacement:
\begin{equation}
\label{eq:LPh}
{\cal L}_{\rm ph}\rho\to\frac{\Gamma_{\rm ph}^{\sigma^{+}a}}{2}{\cal L}[{\sigma}^{+}{a}]\rho
+\frac{\Gamma_{\rm ph}^{a^{\dagger}\sigma^{-}}}{2}{\cal L}[{a}^{\dagger}{\sigma}^{-}]\rho,
\end{equation}
with scattering rates
\begin{equation}
\label{eq:phononrates}
\Gamma_{\rm ph}^{\sigma^{+}a/a^{\dagger}\sigma^{-}} =2g^{\prime2}\,{\rm Re} \left [\int_{0}^{\infty}d\tau\,
e^{\pm i\Delta_{cx} \tau}\! \left (e^{\phi(\tau)}-1 \right )\right],
\end{equation}
where $\Delta_{cx}=\omega_c-\omega_x$ is the cavity-exciton detuning. The replacement follows by making the approximation $X_m(\tau)\approx\exp(-iH_{0}^{\prime}\tau/\hbar){X}_{m}\exp(iH_{0}^{\prime}\tau/\hbar)$, with $H_0' =\hbar\Delta_{xL}{\sigma}^{+}{\sigma}^{-}$, in Eq.~(\ref{eq:FullRates}). This approximation is good when $\eta_x^{-1}$  and  $g^{-1}$ are much smaller than the phonon correlation time (or when the detunings are larger than $\eta_x$ and $g$). With it we capture the dependence of the phonon scattering rates on detuning for detunings that are large enough to have a significant impact on the integral in Eq.~(\ref{eq:phononrates}). In this prescription phonon scattering amounts to a one-way coupling between the driven cavity mode and QD exciton expressed through quantum jumps. There are jumps in two directions---photon creation accompanied by exciton deexitation (rate $\Gamma_{\rm ph}^{a^\dagger\sigma^-}$) and photon annihilation accompanied by excitation of the exciton (rate $\Gamma_{\rm ph}^{\sigma^+a}$)---though Eqs.~(\ref{eq:phase}) and (\ref{eq:phononrates}) yield an asymmetry of rates. The asymmetry of rates allows for phonon-mediated inversion.

For exciton excitation, the Lindblad form of phonon scattering term has two additional contributions:
\begin{align}
\label{eq:LPh2}
{\cal L}_{\rm ph}\rho\to & \frac{\Gamma_{\rm ph}^{\sigma^{+}a}}{2}{\cal L}[{\sigma}^{+}{a}]\rho
+\frac{\Gamma_{\rm ph}^{a^{\dagger}\sigma^{-}}}{2}{\cal L}[{a}^{\dagger}{\sigma}^{-}]\rho +\frac{\Gamma_{\rm ph}^{\sigma^{+}}}{2}{\cal L}[{\sigma}^{+}]\rho
+\frac{\Gamma_{\rm ph}^{\sigma^{-}}}{2}{\cal L}[{\sigma}^{-}]\rho,
\end{align}
with additional scattering rates
\begin{equation}
\label{eq:phononrates2}
\Gamma_{\rm ph}^{\sigma^{+}/\sigma^{-}} =2(\eta_x^\prime)^2\,{\rm Re} \left [\int_{0}^{\infty}d\tau\,
e^{\pm i\Delta_{Lx} \tau}\! \left (e^{\phi(\tau)}-1 \right )\right],
\end{equation}
where $\Gamma_{\rm ph}^{\sigma^+}$ (up scattering) results in incoherent excitation and $\Gamma_{\rm ph}^{\sigma^-}$ (down scattering) causes enhanced decay. Further details appear in Ref.~\cite{Roy&Hughes:PRX11}. The asymmetry in these rates also allows for phonon-mediated inversion.

In the absence of cavity coupling, it is straightforward to solve for the steady-state exciton population analytically from the above ME model. The solution is
\begin{align}
\label{Nx:analytic}
N_x \equiv \braket{\sigma^+\sigma^-}_{\rm ss}= \frac{1}{2} \left [1 + \frac{\Gamma_{\rm ph}^{\sigma^+} - \Gamma_{\rm ph}^{\sigma^-} -\gamma}{\Gamma_{\rm ph}^{\sigma^+} + \Gamma_{\rm ph}^{\sigma^-} + \gamma +  \frac{4(\eta_x^\prime)^2\Gamma_{\rm pol}}{\Gamma_{\rm pol}^2 + \Delta_{Lx}^2} } \right ],
\end{align}
with polarization decay rate $\Gamma_{\rm pol}=\frac{1}{2}(\Gamma_{\rm ph}^{\sigma^+} + \Gamma_{\rm ph}^{\sigma^-} + \gamma +\gamma')$. This formula was recently presented to model photoluminescence experiments on single In(Ga)As QDs (without any connection to inversion) and found to be in excellent agreement with the experiments \cite{StuttgartExps1}. As we show below, the process of phonon-mediated incoherent excitation [$\Gamma_{\rm ph}^{\sigma^+}$ process, shown schematically in Fig.~\ref{fig:schematic2}(a)] can lead to population inversion. Gl\"assl {\em et al.} \cite{Glassl:PRB11} numerically solve for the exciton population in the absence of a cavity, neglecting ZPL decays (neglecting $\gamma$ and $\gamma^\prime$). They find good agreement with an assumed thermal occupation of the dot-photon dressed states, which yields
\begin{align}
\label{Nx:thermal}
N_x^{\rm th} = \frac{1}{2} \left [
1+\frac{ \Delta_{Lx}}{\tilde\Omega}\tanh \left (\frac{\hbar\tilde\Omega}{2k_BT}\right )\right ] ,
\end{align}
where $\tilde\Omega=\sqrt{\Delta_{Lx}^2+4(\eta_x^\prime)^2}$. Below we show that, in comparison to Eq.~(\ref{Nx:analytic}), this formula is only a reasonable approximation for very large pump fields, and even then fails in general by omitting the important influence of the ZPL decays.

The full cavity-QD results which follow are obtained by solving the ME, Eq.~(\ref{ME}), numerically with electron-phonon scattering treated in the Lindblad approximation [Eq.~(\ref{eq:LPh}) or Eq.~(\ref{eq:LPh2})]. As depicted in Fig.~\ref{fig:schematic2}(b)-(d), we adopt a basis of truncated photon states, $\ket{n}$, $n=1,\ldots,N$, and the exciton states $\ket{g}$ and $\ket{e}$. We find that truncation at $N=60$ is required for convergence at the chosen pump levels, especially in the presence of electron-phonon scattering.

For a simplest picture of the inversion mechanism with cavity excitation, Fig.~\ref{fig:schematic2}(d) shows the two harmonic oscillator ladders, coordinated with $\ket{g}$ and $\ket{e}$, that result when the Jaynes-Cummings  interaction term $\hbar  X_g$ is dropped from the polaron-transformed Hamiltonian. With this simplification, if the cavity is detuned to the blue of the zero-photon line, as shown below (Fig.~\ref{fig:2}), $\Gamma_{\rm ph}^{\sigma^+a}$ may dominate over $\Gamma_{\rm ph}^{a^\dagger\sigma^-}$ and pump the exciton into its excited state. For exciton excitation, a similar argument holds based on the asymmetry of $\Gamma_{\rm ph}^{\sigma^+}$ and $\Gamma_{\rm ph}^{\sigma^-}$; it is unclear, however, what the cavity will do, as two quanta excitation can also efficiently load the exciton in the absence of phonon scattering [cf.~Fig.~\ref{fig:schematic2}(c)] \cite{Hughes:PRL11}.

The polaron ME is expected to be valid when the drive rates and exciton-cavity coupling rates are  smaller than the phonon coupling cut-off frequency $\omega_b$ \cite{Nazir:NJP10}, the case of interest in this work. When the drive strength is much larger than $\omega_b$, one can adopt an elegant variational ME approach \cite{Nazir:PRB11}, or resort to numerical path integral or correlation expansion techniques \cite{Glassl:PRB11}. A major advantage of a ME approach is the clear connection to the underlying scattering processes provided and its ability to account for the various decay processes that must typically be included to explain experiments.

\section{Results}
\label{results}
\begin{figure}[t]
\centering
\includegraphics[width=0.75\columnwidth]{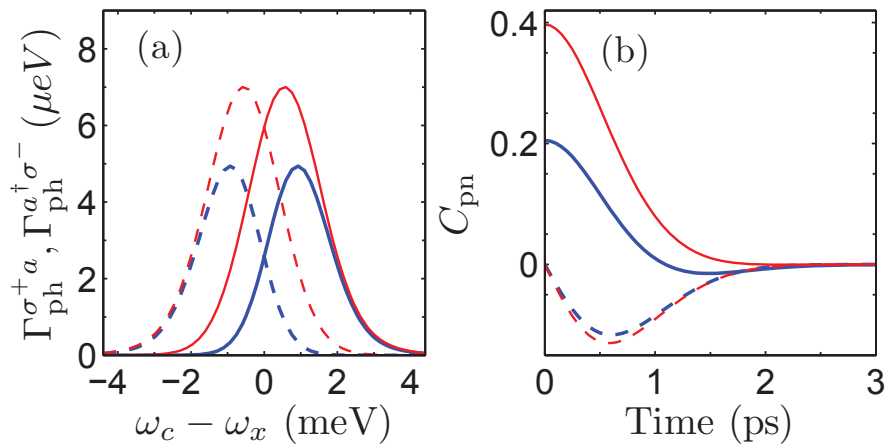}
\vspace{-0.0cm}
\caption{(Color online)
(a) Phonon scattering rates $\Gamma_{\rm ph}^{\sigma^+a}$ (solid) and $\Gamma_{\rm ph}^{a^\dagger\sigma^-}$ (dashed) as a function of cavity detuning from the zero phonon line; for an effective cavity-exciton coupling $g^\prime=0.1\mkern2mu{\rm meV}$, spectral parameters $\omega_b=1\mkern2mu{\rm meV}$, $\alpha_p/(2\pi)^2=0.06\mkern2mu{\rm ps}^2$, and bath temperatures  $T=4\mkern2mu{\rm K}$ (blue, thicker curves) and $T=10\mkern2mu{\rm K}$ (red, thinner curves). The rates $\Gamma_{\rm ph}^{\rm \sigma^+}$ and $\Gamma_{\rm ph}^{\rm \sigma^-}$ corresponding to exciton driving (not shown) have the same functional dependence with $\Delta_{Lx}$ replacing $\Delta_{cx}$ ($\eta_x'^2$ replaces $g'^2$ as the drive). (b)  Corresponding  phonon correlation function, $C_{\rm pn}(t)= \exp[\phi(t)]-1$, showing both the real (solid) and imaginary part (dashed).       }
\label{fig:2}
\end{figure}

We choose material parameters suitable for InAs QDs, with $\omega_b=1\mkern2mu{\rm meV}$ and $\alpha_p/(2\pi)^2=0.06\mkern2mu{\rm ps}^2$ \cite{Roy&Hughes:PRL11,Hughes:PRB11}, and consider a QD-cavity system in the strong coupling regime of cavity QED, with parameters $g^\prime =100\mkern2mu\mu{\rm eV}$, $\gamma=0.5\mkern2mu\mu{\rm eV}$, $\kappa=50\mkern2mu\mu{\rm eV}$, and $\gamma^{\prime}=2\mkern2mu\mu{\rm eV}$.  These numbers are consistent with various QD-cavity experiments and show good agreement with experiments on In(Ga)As QDs, both with \cite{Ulrich:PRL11,Roy&Hughes:PRL11} and without  \cite{StuttgartExps1} cavity coupling. To gain essential insight into the relevant phonon-induced scattering rates, we plot two of them as a function of cavity-exciton detuning, at two different bath temperatures, in Fig.~\ref{fig:2}(a). The figure highlights the asymmetry between $\Gamma_{\rm ph}^{\sigma^+ a}$ and $\Gamma_{\rm ph}^{a^\dagger \sigma^-}$, which is more pronounced at the lower temperature of $4\mkern2mu{\rm K}$ than at $10\mkern2mu{\rm K}$ (when phonon absorption is less likely). Note also that the scattering rates at $10\mkern2mu{\rm K}$ are larger than at $4\mkern2mu{\rm K}$, even though $\braket{B}(10\,{\rm K})=0.84$ is lower than $\braket{B}(4\,{\rm K})=0.91$. We plot the phonon correlation function, $C_{\rm pn}(t)\equiv\exp[\phi(t)]-1$, in Fig.~\ref{fig:2}(b). The plot shows that higher temperatures are more heavily damped, and thus the peak in the scattering rates shifts to lower frequencies at higher temperature. The trend for $\Gamma_{\rm ph}^{\sigma^+}$  and $\Gamma_{\rm ph}^{\sigma^-}$ is exactly the same, only with respect to a $\Delta_{Lx}$ detuning dependence.

\begin{figure}[t]
\centering
\includegraphics[width=0.8\columnwidth]{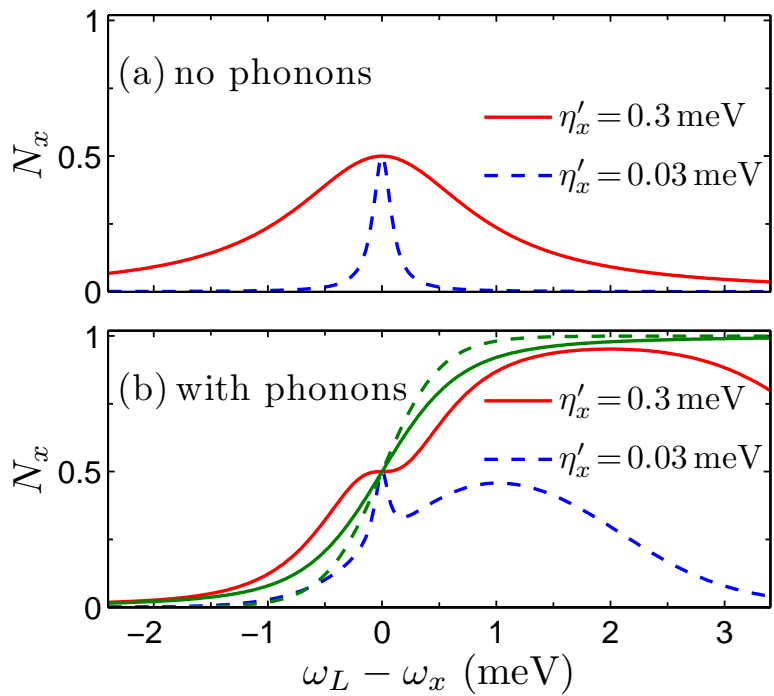}
\vspace{-0.0cm}
\caption{(Color online)
Quantum dot system without cavity coupling at a bath temperature of $4\mkern2mu$K. Steady state exciton populations are plotted for two different drive strengths, $\eta'_x = 0.03\mkern2mu{\rm meV}$ (blue dashed) and $\eta'_x = 0.3\mkern2mu$meV (red solid). (a) Without phonon-induced scattering there is no inversion. (b) With  phonon-induced scattering an inversion ($N_x>0.5$) is produced by the larger drive; the mechanism is incoherent excitation through the phonon bath. We also plot results for the thermal occupation model [Eq.~(\ref{Nx:analytic})] in (b); green dashed curve for $\eta'_x = 0.03\mkern2mu{\rm meV}$ and green solid curve for $\eta'_x = 0.3\mkern2mu{\rm meV}$.}
\label{fig:3n}
\end{figure}

\begin{figure}[t]
\centering
\includegraphics[width=0.8\columnwidth]{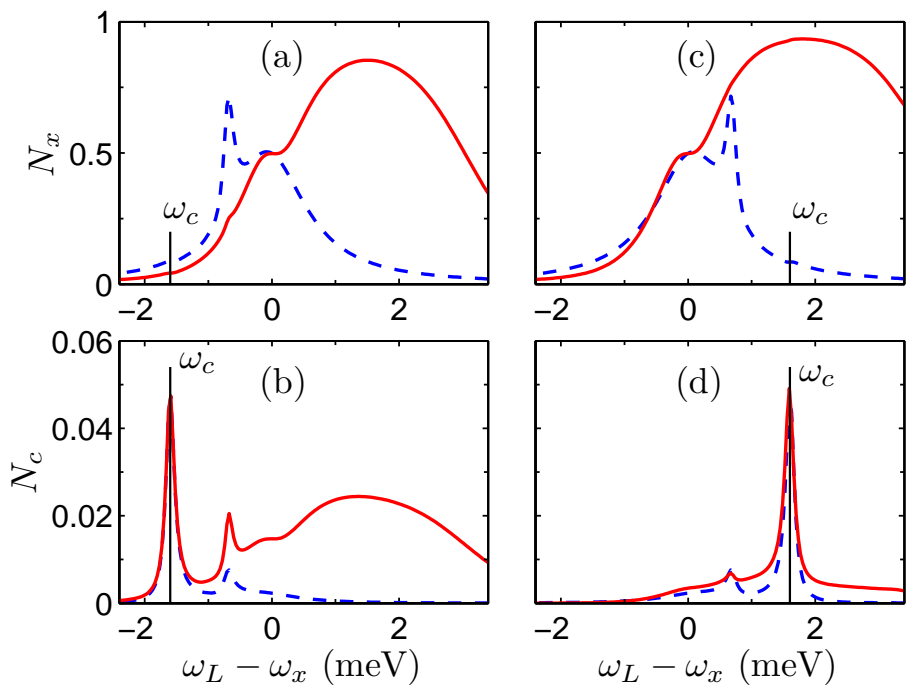}
\vspace{-0.0cm}
\caption{(Color online)
Quantum-dot cavity system with exciton drive $\eta'_x=0.3\mkern2mu{\rm meV}$ and a bath temperature of $4\mkern2mu$K. Steady state exciton and cavity populations are plotted for detunings $\Delta_{cx}=-1.6\mkern2mu{\rm meV}$ (a,b) and $\Delta_{cx}=1.6\mkern2mu{\rm meV}$ (c,d); solid red curves with phonon scattering and blue dashed curves without phonon scattering. The bare cavity resonance is indicated by a vertical black line. The narrow inversion peak appearing without phonon scattering is due to two-photon excitation of the state $\ket{1,e}$ \cite{Savage:PRL88,Hughes:PRL11,Leek:PRB09}.}
\label{fig:4n}
\end{figure}

\begin{figure}[t]
\centering
\includegraphics[width=0.75\columnwidth]{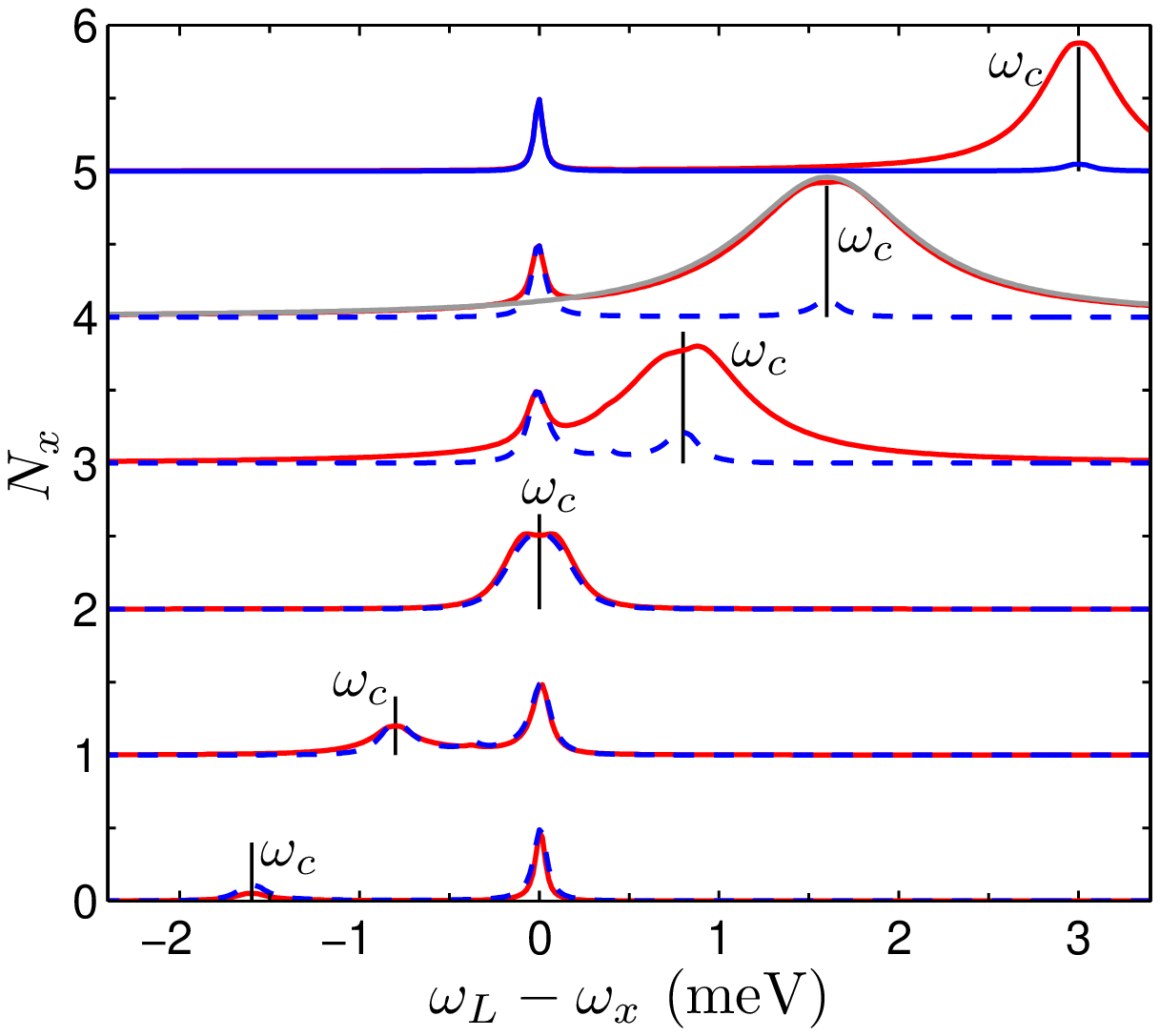}
\vspace{-0.0cm}
\caption{(Color online)
Quantum-dot cavity system with cavity drive $\eta_c=0.3\mkern2mu$meV and a bath temperature of $4\mkern2mu$K. Steady-state exciton populations are plotted as a function of laser frequency for cavity-exciton detunnings $\Delta_{cx}=-1.6$, $-0.8$, $0$, $0.8$, $1.6$, and $3.0\mkern2mu{\rm meV}$ (lower to upper). Blue dashed curves show the population without phonon scattering while the red solid curves include phonon scattering. The bare cavity resonance is indicated by the vertical black line. The grey solid curve (near $\Delta_{cx}=1.6\mkern2mu{\rm meV}$) shows results obtained with the Jaynes-Cummings term omitted from the polaron-transformed Hamiltonian [coupled harmonic ladders, cf.~Fig.~\ref{fig:schematic2}(d)].
}
\label{fig:6}
\end{figure}

With these scattering rates as input, we first explore the steady-state exciton population in the absence of a cavity, i.e., the predictions of the analytical expression Eq.~(\ref{Nx:analytic}). We adopt the lower bath temperature, which yields the larger asymmetry of rates. In Fig.~\ref{fig:3n} we plot the exciton population versus detuning, without [frame (a)] and with [frame (b)] phonon scattering, and for two different strengths of the drive. The larger drive produces substantial power broadening and population inversion when phonon scattering is included. The prediction of inversion is consistent with previously reported results \cite{Tom:PRL05,Glassl:PRB11}. We note that it is obtained here from an analytical expression with a simple physical interpretation, and including ZPL broadening mechanisms that have been shown to be necessary for good agreement with experiments \cite{StuttgartExps1}. To connect with the results of Ref.~\cite{Glassl:PRB11}, we also plot (green curves) the prediction of the thermal occupation model, Eq.~(\ref{Nx:analytic}). It clearly overestimates the populations, and, moreover, fails entirely for lower pump strengths as might be expected.

Next, we add in cavity coupling while keeping the exciton drive. It was recently shown that coupling to an off-resonant cavity in this configuration can invert a simple TLS via two-photon pumping through the state $\ket{1,e}$  [cf.~Fig.~\ref{fig:schematic2}(c)] \cite{Hughes:PRL11}. It is of interest how phonon scattering affects this result. We again adopt the lower bath temperature, and we choose the larger value, $\eta'_x=0.3\mkern2mu {\rm meV}$, for the drive. Figures~\ref{fig:4n}(a) and \ref{fig:4n}(c) show results for cavities red and blue shifted, respectively, by $1.6\mkern2mu {\rm meV}$;  corresponding mean cavity photon numbers appear in Fig.~\ref{fig:4n}(b,d). The chosen cavity-exciton detuning realizes inversion via two-photon resonance  while also having a large asymmetry of the phonon rates. In the absence of phonon scattering (blue dashed lines), we confirm the results of Ref.~\cite{Hughes:PRL11}: an inversion peak appears approximately midway between the cavity and exciton resonances. When phonon scattering is included, the peak is either suppressed (red shifted cavity) or subsumed by an enhanced domain of inversion associated with the $\Gamma_{\rm ph}^{\sigma^+}$ process (blue shifted cavity). As might be anticipated from Eq.~(\ref{Nx:analytic}) and Ref.~\cite{Glassl:PRB11}, both cavity detunings show substantial inversion due to the $\Gamma_{\rm ph}^{\sigma^+}$ process on the blue side of the exciton; although, details of the $N_x$ and $N_c$ lineshapes depend on cavity detuning, with the $N_c$ profiles, in particular, retaining clear signatures of the two-photon resonance.

It remains to explore cavity coupling with cavity excitation. We again compare exciton populations with phonon scattering against those with the phonons turned off (note that $g$ is replaced by $g^\prime$ in the latter case). The comparison is made in Fig.~\ref{fig:6}, which shows the exciton population passing through a peak as a function of $\Delta_{cx}$, more or less in line with the peak in the scattering rates. Most notably, large population inversions are obtained when the cavity is blue-shifted, the configuration that allows phonon scattering between ground- and excited-state branches of the photon ladder of states to excite the exciton [cf.~Fig.~\ref{fig:schematic2}(d)]. The reverse process, \emph{cavity feeding}, has been identified in semiconductor cavity-QED studies \cite{Calic:PRL11,Hughes:PRB11,HohenesterPRB:2010,Roy&Hughes:PRL11}. Here we see that the phonon-mediated $\Gamma_{\rm ph}^{\sigma^+a}$ process can create large population inversions. We stress that the mechanism is quite different to the one reported in Refs.~\cite{Savage:PRL88} and \cite{Hughes:PRL11} and is considerably more efficient [cf.~Fig.~\ref{fig:4n}(a,c), dashed blue lines]. Not only do we see pronounced inversion in the presence of phonon scattering, with $N_x>0.9$ at $\Delta_{cx}=1.6\mkern2mu{\rm meV}$ in Fig.~\ref{fig:6}, but also significant inversion over a broad detuning range. We see, however, that the role of the cavity resonance is much more pronounced than with the exciton drive. For cavity excitation the $\Gamma_{\rm ph}^{\sigma^+a}$ process dominates, while for exciton excitation the $\Gamma_{\rm ph}^{\sigma^+}$ process dominates. Although this inverse of cavity-feeding is to be expected, it is frequently omitted from QD ME theories and, to the best of our knowledge, has not been noticed in any QD cavity-QED experiments to date. Phonon-mediated population inversion is unique to the solid state environment where it may allow for cavity-pumped single exciton lasing.  It is important to note that all these results are reproduced by the full non-Lindblad ME \cite{Roy&Hughes:PRX11}.

At the optimal cavity-exciton detuning, near $\Delta_{cx}=1.6\mkern2mu{\rm meV}$, the exciton population is greater than $0.9$, and even higher numbers result at higher values of the drive. As indicated above [cf.~Fig.~\ref{fig:schematic2}(d)], the mechanism underlying this cavity-excited inversion is seen, in a simplest model, to be electron-phonon scattering between a pair of photon ladders, one coordinated with the ground state of the exciton and the other with the excited state. This simplest model neglects the Jaynes-Cummings interaction term $\hbar  X_g$ in the polaron-transformed Hamiltonian [Eq.~(\ref{eq:H_S})]. To test it, we also carried out calculations with the Jaynes-Cummings term omitted. Results for $\Delta_{cx}=1.6\mkern2mu{\rm meV}$ appear as the solid grey curve in Fig.~\ref{fig:6}. They confirm the qualitative correctness of the model; although, as one would expect, the exciton resonance is completely missed.

\begin{figure}[t]
\centering
\includegraphics[width=0.9\columnwidth]{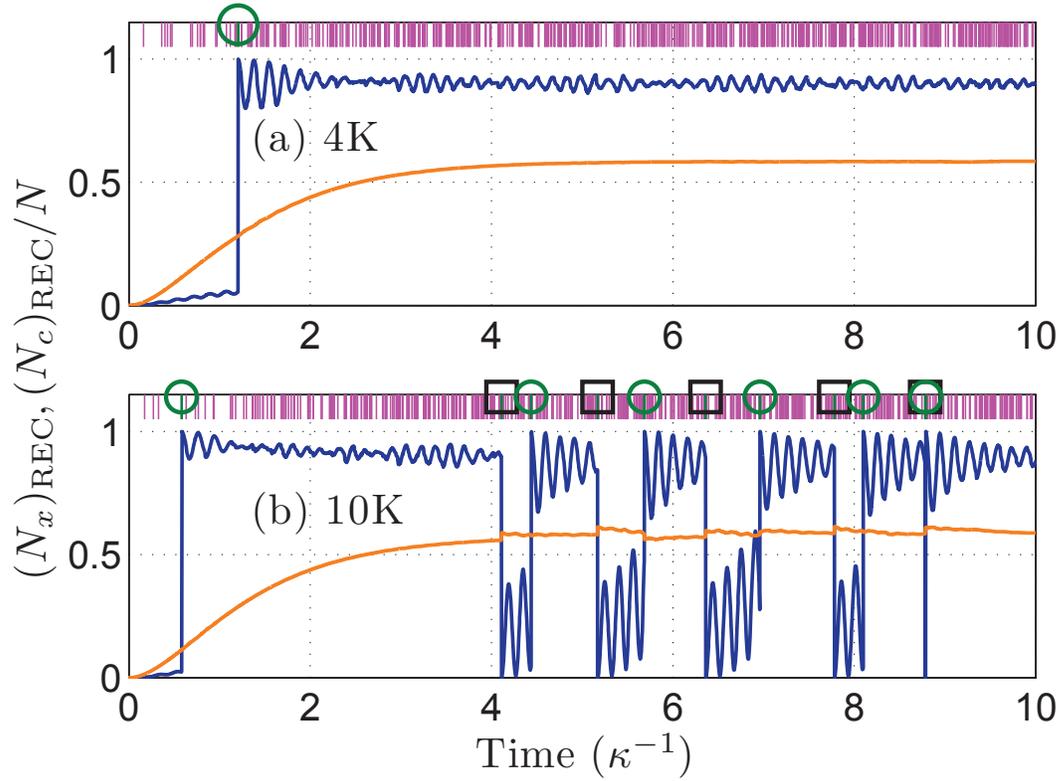}
\vspace{-0.0cm}
\caption{(Color online)
Sample quantum trajectories at (a) $4\mkern2mu{\rm K}$ and (b) $10\mkern2mu{\rm K}$, for optimal detuning, $\Delta_{cx}=1.6\mkern2mu{\rm meV}$, and other parameters as in Fig.~\ref{fig:2}. Orange (light) lines display the photon number expectation $(N_c)_{\rm REC}/N=\langle a^\dagger a\rangle_{\rm REC}/N$ ($N=60$) and blue (dark) lines the exciton expectation $(N_x)_{\rm REC}=\langle\sigma^+\sigma^-\rangle_{\rm REC}$. The expectations are {\em conditioned} upon the record of quantum jumps: either photon decay (magenta lines) or phonon-mediated scattering [green circles (black squares) for scattering at rate $\Gamma_{\rm ph}^{\sigma^+a}$ ($\Gamma_{\rm ph}^{a^\dagger\sigma^-}$)].}
\label{fig:7}
\end{figure}

The quantum trajectory simulations \cite{qt} presented in Fig.~\ref{fig:7} provide insight into the role of the neglected Jaynes-Cummings term.  For a detuning of $\Delta_{cx}=1.6~$meV, the phonon scattering dynamic  at $4\mkern2mu{\rm K}$ and $10\mkern2mu{\rm K}$ is displayed and compared.  Short magenta lines signal cavity photon jumps, while green circles and black squares identify phonon scattering jumps---photon annihilation accompanied by exciton exitation (green circles) and photon creation accompanied by deexcitation (black squares). Phonon scattering excites the exciton within a cavity lifetime. The excitation is maintained over a relatively long time at $4\mkern2mu{\rm K}$, and eventually lost primarily through radiative decay (the $\gamma$-jump is not seen in the figure). Reverse phonon scattering is the main source of deexcitation at $10\mkern2mu{\rm K}$, resulting in a reduced population (from $0.9$
 to $0.8$). The Jaynes Cummings term produces the oscillation between photon scattering events. It hardly alters the pattern of quantum jumps but reduces the mean population a little.

\begin{figure}[t]
\centering
\includegraphics[width=0.75\columnwidth]{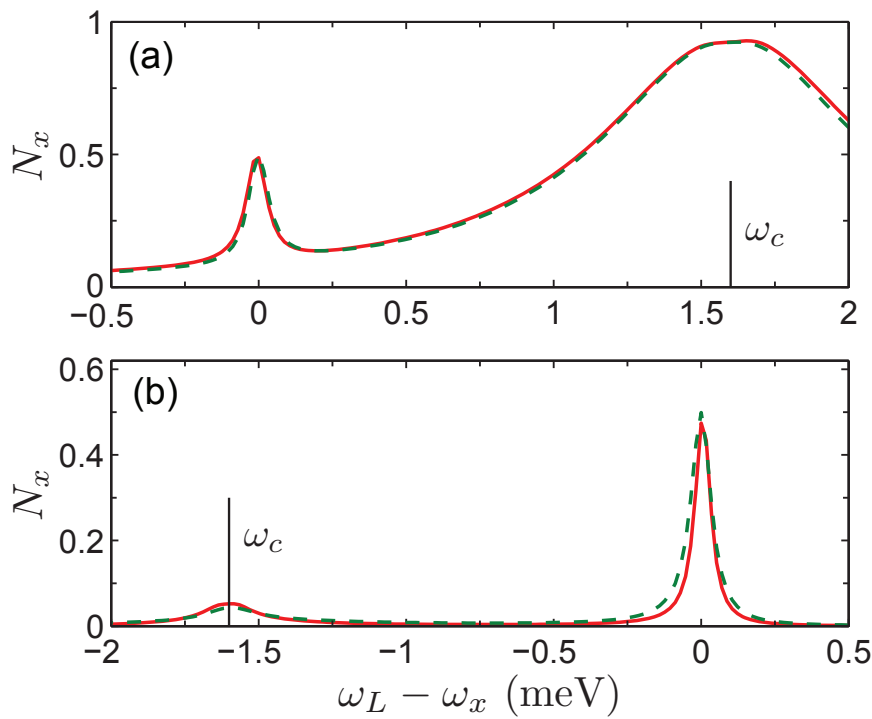}
\vspace{-0.0cm}
\caption{(Color online)
Same as Fig.~\ref{fig:6}, for detunings $\Delta_{cx}=\pm 1.6~$meV. The numerical solution of the ME (red solid curves) is compared with  Eq.~(\ref{Nx:analytic2}) (green dashed curves). To improve the fit the decay rate enhancement has been set at $p=2.5$.
}
\label{fig:6Extra}
\end{figure}

Finally, it is instructive to develop a formula similar to Eq.~(\ref{Nx:analytic}) to help explain the qualitative difference in the response for cavity versus exciton driving. To this end, we first ignore the exciton-cavity coupling, so the cavity simply fills up with coherent light. The steady-state density operator factorizes as $\rho=\rho_x\ket{\alpha}\bra{\alpha}$, with coherent state amplitude
\begin{align}
\alpha = \frac{\eta_c}{\kappa-i\Delta_{Lc}}.
\end{align}
We then substitute the \emph{ansatz} into the ME, take the trace over the cavity mode, and make the approximation $\braket{\alpha|aa^\dagger|\alpha}\approx|\alpha|^2$, i.e., write $aa^\dagger=a^\dagger a+1\approx a^\dagger a$. This yields a ME with effective exciton drive $g^\prime\alpha$. The solution for the exciton population takes the same form as Eq.~(\ref{Nx:analytic}):
\begin{align}
\label{Nx:analytic2}
N_x = \frac{1}{2} \left [1 + \frac{\tilde\Gamma_{\rm ph}^{\sigma^+} - \tilde\Gamma_{\rm ph}^{\sigma^-} -\tilde\gamma}{\tilde\Gamma_{\rm ph}^{\sigma^+} + \tilde\Gamma_{\rm ph}^{\sigma^-} + \tilde\gamma +  \frac{4(\eta_x^\prime)^2\tilde\Gamma_{\rm pol}}{\tilde\Gamma_{\rm pol}^2 + \Delta_{Lx}^2} } \right ],
\end{align}
with $\tilde\Gamma_{\rm ph}^{\sigma^+}=|\alpha|^2\Gamma_{\rm ph}^{\sigma^+a}$,
$\tilde\Gamma_{\rm ph}^{\sigma^-}=|\alpha|^2\Gamma_{\rm ph}^{a^\dagger \sigma^-}$,
$\eta_x^\prime = g'\eta_c[\tilde\kappa^2+\Delta_{Lx}^2]^{-1/2}$, and
$\tilde\Gamma_{\rm pol}=\frac{1}{2}(\tilde\Gamma_{\rm ph}^{\sigma^+}+\tilde\Gamma_{\rm ph}^{\sigma^-}+\tilde\gamma+\gamma')$. In place of $\kappa$ and $\gamma$, enhanced rates $\tilde\kappa=p\kappa$ and $\tilde\gamma=p\gamma$ have been introduced. The fitting parameter $p$ aims to account for absorption (enhancing $\kappa$) and the Purcell effect (enhancing $\gamma$), which are overlooked when the exciton-cavity coupling is ignored. Figure~\ref{fig:6Extra} compares the full numerical solution (red solid curve) with Eq.~(\ref{Nx:analytic2}) (green dashed curves) for two different values of cavity-exciton detuning, and for the choice $p=2.5$. The agreement is surprisingly good for both positive [Fig.~\ref{fig:6Extra}(a)] and negative [Fig.~\ref{fig:6Extra}(b)] detunings, and, in contrast to the grey curve in Fig.~\ref{fig:6}, the model captures the exciton peak as well as the peak around the cavity resonance. The model demonstrates the effect of cavity filtering on the phonon-mediated scattering rates $\tilde\Gamma_{\rm ph}^{\sigma^+}$ and $\tilde\Gamma_{\rm ph}^{\sigma^-}$, which are both proportional to $|\alpha|^2$. This filtering is important in the cavity-driven system, as it is the primary control over when the inversion turns off as the drive is detuned. The situation is quite different with direct driving of the exciton and no cavity. In that case the inversion process turns off because of the $\Delta_{Lx}$ factor inside the integral, Eq.~(\ref{eq:phononrates2}), defining the phonon scattering rates.

\section{Conclusions}
\label{conclusions}
We have reported on a number of different ways to create exciton inversion in a semiconductor quantum-dot cavity system with cw drive. The underlying mechanism exploits a highly efficient phonon-mediated scattering process to create excitons via direct incoherent excitation or through the annihilation of a cavity photon; the reverse process takes place at a much reduced rate in a low temperature bath. At a phonon bath temperature of $4\mkern2mu{\rm K}$, exciton populations greater than 0.9 are easily achievable. Direct exciton excitation and excitation through a coupled cavity are found to be qualitatively different. In particular, in the latter case, the effect of cavity filtering is important and depends upon the sign of the cavity-exciton detuning.

\section*{Acknowledgments}

This work was supported by the National Sciences and Engineering Research Council of Canada and the Marsden Fund of the Royal Society of New Zealand. The authors thank P. Michler, S. Ulrich, A. Ulhaq, and S. Weiler for useful discussions.

\end{document}